# Non-Topological Approach to the Stability of 't Hooft-Polyakov Monopole

**K. Rasem Qandalji**
*Amer Institute, P.O.Box 1386, Sweileh, 11910, JORDAN*

**Abstract.** At classical level, dynamical derivation of the properties and conservation laws for topologically non-trivial systems from Noether theorem versus the derivation of the system's properties on topological grounds are considered as distinct. We do celebrate any agreements in results derived from these two distinct approaches: i.e. the dynamical versus the topological approach. Here we consider the Corrigan-Olive-Fairlie-Nuyts solution based on which we study the stability of the 't Hooft- Polyakov outer field, known as its Higgs vacuum, and derive its stability, dynamically, from the equations of motion rather than from the familiar topological approach. Then we use our derived result of the preservation of the Higgs vacuum asymptotically to derive the stability of the 't Hooft-Polyakov monopole, even if inner core is perturbed, where we base that on observing that the magnetic charge must be conserved if the Higgs vacuum is preserved asymptotically. We also, alternatively, note stability of 't Hooft-Polyakov monopole and the conservation of its magnetic charge by again using the result of the Higgs vacuum asymptotic preservation to use Eq.(5) to show that no non-Abelian radiation allowed out of the core as long as the Higgs vacuum is preserved and restored, by the equations of motion, if perturbed. We start by deriving the asymptotic equations of motion that are valid for the monopole's field outside its core; next we derive certain constraints from the asymptotic equations of motion of the Corrigan-Olive-Fairlie-Nuyts solution to the 't Hooft-Polyakov monopole using the Lagrangian formalism of singular theories, in particular that of Gitman and Tyutin. The derived constraints will show clearly the stability of the monopole's Higgs vacuum its restoration by the equations of motion of the Higgs vacuum, if disturbed.

## 1. Introduction and Motivation.

For systems with topologically non-trivial field configurations and degenerate vacua: conservation laws, conserved currents and conserved quantum charges are usually derived independently of the equations of motion. Here, the conserved laws and charges are topological, and, at the classical level, they don't follow from the equations of motion, i.e. they are not derived from the Noether theorem which is based on the symmetry of the system's given Lagrangian. Only at the "quantum" level, and verified only in the simple case of Sine-Gordon (1+1) model versus the Thirring model [1], the equivalence of both symmetry currents was explicitly demonstrated [2,3]. Extending these findings to the 't Hooft -Polyakov monopole is more subtle. Still at the classical level, the system's properties and characteristics, the conservation laws, currents, and charges derived from the topological versus the Noether approaches are considered as distinct; and hence any agreements in results from both approaches should be celebrated.

In the work here, we consider, as one of the system's conserved feature, the "stability" of the 't Hooft-Polyakov monopole's outer field, i.e. the system's Higgs vacuum region, against perturbations. We show that versus what is known about the stability of the system's outer field, or its Higgs vacuum, derived, usually, on topological grounds, [4,5,6]; stability of the system's field can also be, distinctly and equivalently, derived, as will be done here, from the equations of motion in that region based on the Carrigan *et al* general solution [12].

For the system at hand, i.e., 't Hooft-Polyakov monopole, to be physically acceptable; its energy must be finite at all time. The energy finiteness should not be violated as the system develops in time based on the system's equations of motion. To insure the system's energy finiteness, we are led to the requirement [7,8] that, except for a localized finite region where the monopole's core resides, the outer region must, at all time, satisfy the Higgs vacuum conditions, given in Eqs. (2, 3), below.

The physicality requirement of the model's energy finiteness is what causes the existence of the 't Hooft-Polyakov magnetic monopole: For any solution of the Georgi-Glashow model [9], energy finitness of the model requires the existence of long-range magnetic field unless the vacuum is topologically trivial, for review see[10,11]; but if the degenerate vacuum is realized in a topologically non-trivial way, then, there exists no non-singular gauge transformation that can make the Higgs field asymptotically constant, and hence the energy integral will diverge unless the squared gradient of the Higgs field term in the energy integral is replaced by the square of the covariant derivative of the Higgs field; which means: allowing for local independent gauge transformations. In this case, the space components of the gauge potential cannot, in general, vanish. Space components of the gauge potentials should decrease as inverse of the distance from the origin, so that they can be adjusted to cancel out any divergent terms, in the energy integral, due to the gradient of the Higgs field. Since the space components of gauge potentials decrease as inverse of distance from the monopole's core; then that will result in space-space magnetic components of the gauge field strengths reducing as inverse squared of distance and hence, make up for the increase in surface area with distance and, therefore, results in constant total flux of magnetic fields; i.e., a long range magnetic field associated with the single surviving generator that annihilates the vacuum asymptotically.

## 2. Preliminaries. Higgs vacuum in the Corrigan-Olive-Fairlie-Nuyts solution.

Outside the interior of the 't Hooft-Polyakov monopole; we have a long range magnetic field that looks like Dirac's monopole field. The non-singular core of the 't Hooft-Polyakov monopole is based on the Georgi-Glashow model[3], which consists of SO(3) gauge fields interacting with Lorentz-scalar Higgs isotriplet, $\boldsymbol{\phi} = (\phi_1, \phi_2, \phi_3)$. In this model; the SO(3) gauge symmetry group, generated by $T_a$'s, is broken spontaneously, by the degenerate non-vanishing vacuum, down to U(1) gauge group that is generated by the charge, $\boldsymbol{\phi} \cdot \mathbf{T}/a$. This single surviving unbroken generator varies with, and also, annihilates the Higgs field asymptotically outside the monopole's core. A massless gauge potential is associated with this single unbroken generator; this gauge potential can be identified outside the monopole's core with the electro-magnetic field.

The Lagrangian of the Georgi-Glashow model is,

$$\mathcal{L} = -\frac{1}{4} G_a^{\mu\nu} G_{\mu\nu a} + \frac{1}{2} \mathcal{D}^\mu \boldsymbol{\phi} \cdot \mathcal{D}_\mu \boldsymbol{\phi} - V(\boldsymbol{\phi}) , \tag{1}$$

where we use metric with signature $= -2$, and where, $V(\boldsymbol{\phi}) = \frac{1}{4}\lambda(\phi_1^2 + \phi_2^2 + \phi_3^2 - a^2)^2$, and where $W_a^\mu$ is the gauge potential. The gauge field strength, $G_a^{\mu\nu}$, is given by: $G_a^{\mu\nu} = \partial^\mu W_a^\nu - \partial^\nu W_a^\mu - e\varepsilon_{abc} W_b^\mu W_c^\nu$.

Gauge fields associated with the two broken generators will *eat up* the two Goldstone bosons available and acquire mass and hence will be short range. The size of HP monopole's interior is estimated using the Compton wavelength of the massive gauge bosons associated with the two broken generators and the mass of the surviving Higgs particle.

Again; the model's energy finiteness implies that, asymptotically, the Higgs vacuum conditions must be satisfied, i.e., both Eqs. (2, 3) below:

$$\mathcal{D}^\mu \boldsymbol{\phi} \equiv \partial^\mu \boldsymbol{\phi} - e\mathbf{W}^\mu \wedge \boldsymbol{\phi} = 0 ; \tag{2}$$

$$\text{and, } \phi_1^2 + \phi_2^2 + \phi_3^2 - a^2 = 0 \ (\Rightarrow V(\boldsymbol{\phi}) = 0) , \tag{3}$$

or, in Higgs vacuum: $\boldsymbol{\phi} \cdot \delta\boldsymbol{\phi} = 0$, or equivalently, $\phi_a \partial^\mu \phi_a = 0, \ (\forall \mu = 0,...,3)$ (3.a)

The most general form of $\mathbf{W}^\mu$ in Higgs vacuum is (Corrigan *et al*, [12])

$$\mathbf{W}^\mu = \frac{1}{a^2 e} \boldsymbol{\phi} \wedge \partial^\mu \boldsymbol{\phi} + \frac{1}{a} \boldsymbol{\phi} A^\mu , \tag{4}$$

where $A^\mu$ is arbitrary. It follows that:

$$\mathbf{G}^{\mu\nu} = \frac{1}{a} \boldsymbol{\phi} F^{\mu\nu} ; \tag{5}$$

where,

$$F^{\mu\nu} = \frac{1}{a^3 e} \boldsymbol{\phi} \cdot (\partial^\mu \boldsymbol{\phi} \wedge \partial^\nu \boldsymbol{\phi}) + \partial^\mu A^\nu - \partial^\nu A^\mu . \tag{6}$$

Also note, *for future use below*, that the monopole's magnetic charge is evaluated [,] by the integral over a surface, $\Sigma$, that is completely in the Higgs vacuum, where Eqs.(2, 3) are satisfied, and surrounds the monopoles core where Eqs. (2, 3) fail, [,]:

$$g = \int_\Sigma \mathbf{B} \cdot d\mathbf{S} = \frac{-1}{2a^3 e} \int_\Sigma \varepsilon_{ijk} \boldsymbol{\phi} \cdot (\partial^j \boldsymbol{\phi} \wedge \partial^k \boldsymbol{\phi}) \, dS^i \,, \tag{7}$$

where, again as shown in [,], the monopole's magnetic charge given by Eq(7) above is invariant under continuous nonsingular time development of the Higgs field, continuous gauge transformation and any continuous regular altering of the surface of integral, $\Sigma$, provided it remains in the Higgs vacuum.

For the Lagrangian of Higgs vacuum, and using Eqs. (2,3,5,6) above, it will reduce there to:

$$\mathcal{L} = -\frac{1}{4} G_a^{\mu\nu} G_{\mu\nu a} \,;$$

and on account of Eqs.(5, 6) we get:

In the Higgs vacuum the Lagrangian reduces to

$$\begin{aligned}\mathcal{L} &= -\frac{1}{4} F^{\mu\nu} F_{\mu\nu} \\ &= \frac{-\varepsilon_{ijk}\varepsilon_{rst}}{4a^6 e^2} \phi_i \phi_r \partial^\mu \phi_j \partial^\nu \phi_k \partial_\mu \phi_s \partial_\nu \phi_t - \frac{1}{2}\left(\partial^\mu A^\nu - \partial^\nu A^\mu\right)\partial_\mu A_\nu - \frac{\varepsilon_{ijk}}{a^3 e} \phi_i \partial^\mu \phi_j \partial^\nu \phi_k \partial_\mu A_\nu\end{aligned} \tag{8}$$

In this work we will verify that stability of the Higgs vacuum, defined below, not in the sense that topologically non-equivalent classes of Higgs field configurations are separated by infinite energy barriers; but we will, rather, verify the stability of the vacuum in the Corrigan-Olive-Fairlie-Nuyts solution,[12], by showing that the equations of motion derived from this solution will restore the Higgs vacuum asymptotically against perturbations and hence will guarantee the system's energy finiteness at all time.

## 3. Constraints from the Model's Lagrangian Dynamics.

In the Lagrangian formalism, see Gitman and Tyutin [13,14], of singular physical systems; the equations of motion and the constraints based on them can be derived from noting that the gauge freedom of these singular systems implies that, the Euler-Lagrange equations of motion of the fields, of the theory at hand, are not all independent; i.e., we should be able to construct identities out of these equations of motions. These identities should be satisfied by combinations of some rows (columns) of the Hessian matrix too; where the Hessian matrix, $\mathcal{M}$, is defined as the second derivative of the *local* Lagrangian density, with respect to the velocities of all the fields available in the theory, ($\varphi_a, \varphi_b, etc...$): $\mathcal{M} \equiv \left\| \mathcal{M}_{ab} = \dfrac{\partial^2 \mathcal{L}}{\partial \dot{\varphi}_a \partial \dot{\varphi}_b} \right\|$.

Further independent combinations of some rows (columns) of the Hessian matrix might also satisfy even more identities than those satisfied by the equations of motion, and that should result in putting constraints on the fields available in the theory and their velocities. These constraints should always be satisfied on any *genuine* path allowed by the equations of motion.

The Euler-Lagrange Equations of motion of our system in Higgs vacuum, given by Eqs.(2-8), are:

$$0 = \frac{\delta S}{\delta \phi_m(x)} \equiv \frac{\partial \mathcal{L}}{\partial \phi_m(x)} - \partial^\sigma \frac{\partial \mathcal{L}}{\partial \partial^\sigma \phi_m(x)} = \frac{\varepsilon_{mjk}}{a^3 e}\left( \phi_j \partial^\mu \phi_k \partial^\nu F_{\mu\nu} - \frac{3}{2} F_{\mu\nu} \partial^\mu \phi_j \partial^\nu \phi_k \right),$$

$$0 = \frac{\delta S}{\delta A^\nu(x)} \equiv \frac{\partial \mathcal{L}}{\partial A^\nu(x)} - \partial^\mu \frac{\partial \mathcal{L}}{\partial \partial^\mu A^\nu(x)} = \partial^\mu F_{\mu\nu}.$$

(9)

Equations of motion, Eqs.(9), can be re-arranged [13,14,15] in order of time derivatives of the fields: i.e., the second time derivatives of the fields multiplying the elements of the Hessian matrix and the rest of the terms of lower time derivatives grouped in, call them, $\mathcal{K}_{\phi_m}$'s, and $\mathcal{K}_{A^\nu}$'s:

$$\left.\begin{array}{l} 0 = \dfrac{\delta S}{\delta \phi_m(x)} \equiv \mathcal{K}_{\phi_m}(\phi_l, \partial^\sigma \phi_n, A^\eta, \partial^\nu A^\mu) - \mathcal{M}_{\phi_m \phi_h}\ddot{\phi}_h - \mathcal{M}_{\phi_m A^\mu}\ddot{A}^\mu \\[6pt] 0 = \dfrac{\delta S}{\delta A^\nu(x)} \equiv \mathcal{K}_{A^\nu}(\phi_l, \partial^\sigma \phi_n, A^\eta, \partial^\nu A^\mu) - \mathcal{M}_{A^\nu \phi_h}\ddot{\phi}_h - \mathcal{M}_{A^\nu A^\mu}\ddot{A}^\mu \end{array}\right\}$$

(10)

where,

$$\mathcal{K}_{\phi_m} = \frac{\varepsilon_{mjk}}{a^3 e}\left[ -\frac{3}{2} F_{\mu\nu} \partial^\mu \phi_j \partial^\nu \phi_k + \phi_j \partial^\mu \phi_k \partial^i F_{\mu i} + \phi_j \partial^i \phi_k \left( \partial_i \dot{A}_0 - \frac{\varepsilon_{rst}}{a^3 e} \phi_r \dot{\phi}_s \partial_i \dot{\phi}_t \right) \right];$$

$$\mathcal{K}_{A^0} = \partial^i F_{i0};$$

$$\mathcal{K}_{A^j} = \partial^i F_{ij} - \partial_j \dot{A}_0 + \frac{\varepsilon_{rst}}{a^3 e} \phi_r \dot{\phi}_s \partial_j \dot{\phi}_t;$$

(11)

and, using Eq.(8), the elements of the *symmetric* Hessian matrix, $\mathcal{M}$, are:

$$\mathcal{M}_{\phi_l \phi_h} = \frac{\partial^2 \mathcal{L}}{\partial \dot{\phi}_l \partial \dot{\phi}_h} = -\frac{\varepsilon_{lmn}\varepsilon_{hrt}}{a^6 e^2} \phi_m \phi_r \partial^k \phi_n \partial_k \phi_t; \qquad \mathcal{M}_{\phi_l A^j} = \frac{\partial^2 \mathcal{L}}{\partial \dot{\phi}_l \partial \dot{A}^j} = \frac{\varepsilon_{lmn}}{a^3 e} \phi_m \partial_j \phi_n;$$

$$\mathcal{M}_{\phi_l A^0} = \frac{\partial^2 \mathcal{L}}{\partial \dot{\phi}_l \partial \dot{A}^0} = 0; \qquad \mathcal{M}_{A^\mu A^0} = \frac{\partial^2 \mathcal{L}}{\partial \dot{A}^\mu \partial \dot{A}^0} = 0; \qquad \mathcal{M}_{A^i A^j} = \frac{\partial^2 \mathcal{L}}{\partial \dot{A}^i \partial \dot{A}^j} = -g_{ij}.$$

(12)

From the above Hessian matrix, Eqs.(12), and in addition to others, we get the following independent identities of interest to us; where summation over repeated indices is understood:

$$\left.\begin{array}{l}\dfrac{\partial^2 \mathcal{L}}{\partial \dot{\phi}_l \partial \dot{\phi}_h}+\left(\dfrac{\varepsilon_{lmn}}{a^3 e}\phi_m \partial^k \phi_n\right)\dfrac{\partial^2 \mathcal{L}}{\partial \dot{A}^k \partial \dot{\phi}_h}\equiv 0,\\[2mm] \dfrac{\partial^2 \mathcal{L}}{\partial \dot{\phi}_l \partial \dot{A}^j}+\left(\dfrac{\varepsilon_{lmn}}{a^3 e}\phi_m \partial^k \phi_n\right)\dfrac{\partial^2 \mathcal{L}}{\partial \dot{A}^k \partial \dot{A}^j}\equiv 0,\\[2mm] for\ l=1,2,3;\ and\ \forall h\,(=1,2,3);\forall j\,(=1,2,3)\end{array}\right\} ; \quad (13)$$

Directly off Eqs.(13), we read the elements of the three independent zero-eigenvectors of the Hessian matrix, call them, $u^{\phi_m}{}_{(l)}$, $u^{A^\mu}{}_{(l)}$; ($l=1,2,3$).

*Symbolically*; the identities are written as: $u^a{}_{(l)}\mathcal{M}_{ab}\equiv 0$:

$$u^{\phi_m}{}_{(l)}=\delta_{ml}; \quad u^{A^k}{}_{(l)}=\dfrac{\varepsilon_{lmn}}{a^3 e}\phi_m \partial^k \phi_n; \quad u^{A^0}{}_{(l)}\ \text{is arbitrary},\quad (for\ l=1,2,3). \quad (14)$$

If we multiply the equations of motion, Eq.(10), by the vectors $u_{(l)}$, Eqs.(14), we will, then, identically eliminate the second time derivative part in Eqs.(10); So, on genuine trajectories, i.e. where equations of motion are satisfied, Eqs.(10,13,14) lead to:

$$0=u^a_{(l)}\mathcal{K}_a=\delta_{lm}\mathcal{K}_{\phi_m}+\left[\dfrac{\varepsilon_{lmn}}{a^3 e}\phi_m \partial^k \phi_n\right]\mathcal{K}_{A^k}+u^{A^0}{}_{(l)}\mathcal{K}_{A^0}, \quad (for,\ l=1,2,3.) \quad (15)$$

Since $u^{A^0}{}_{(l=1,2,3)}$ is arbitrary, [see Eq.(14)], we, purposely, pick: $u^{A^0}{}_{(l=1,2,3)}=\dfrac{\varepsilon_{lmn}}{a^3 e}\phi_m \partial^0 \phi_n$, and using Eqs.(11), then, on genuine trajectories: Eq.(15) reduces to

$$0=u^a_{(l)}\mathcal{K}_a=-\left(\dfrac{3}{2}\right)\dfrac{\varepsilon_{ljk}}{a^3 e}F_{\mu\nu}\partial^\mu \phi_j \partial^\nu \phi_k, \quad (for,\ l=1,2,3.) \quad (16)$$

We form the following linear combinations of Eqs.(16), or in other words, we form new vectors, $v^a_{(k)}$ and $w^a$, from the original $u^a_{(l)}$'s and that will result in two independent identities given by Eqs.(17) below, where these will be as shown to be identities on account of Eq.(3.a); as well as one constraint, given by Eq.(18) below, that is independent of the two identities of Eqs.(17), and it has to vanish, due to equations of motion, on genuine trajectories of the system.

Eqs.(16) will, now, be, equivalently, replaced by the two identities of Eqs.(17), and the constraint in Eq.(18):

$$0=v^a_{(m)}\mathcal{K}_a=\varepsilon_{mil}\phi_i u^a_{(l)}\mathcal{K}_a=-\varepsilon_{mil}\varepsilon_{ljk}\left(\dfrac{3}{2a^3 e}\right)F_{\mu\nu}\phi_i \partial^\mu \phi_j \partial^\nu \phi_k, \quad (for,\ m=1,2) \quad (17)$$

where, $v^a_{(m)}\equiv \varepsilon_{mil}\phi_i u^a_{(l)}$. We also have the constraint:

$$0=w^a \mathcal{K}_a, \qquad . \qquad (18)$$

where, $w^a = \phi_l u^a_{(l)}$.

## 4. Stability of the Outer Field (Higgs Vacuum) from Equations of Motion.

Recall, first, that, the equation, $\boldsymbol{\phi} \cdot \delta \boldsymbol{\phi} = 0$, Eq.(3.a), is true in the Higgs vacuum region, since the Higgs field, $\boldsymbol{\phi}$, belongs, there, to the vacuum manifold, $\mathcal{M}_0$, which is defined as: $\mathcal{M}_0 \equiv \left\{ \boldsymbol{\phi} : V(\boldsymbol{\phi}) = |\boldsymbol{\phi}|^2 - a^2 = 0 \right\}$.

Eqs.(17) are identities on account of Eq.(3.a): $\boldsymbol{\phi} \cdot \delta \boldsymbol{\phi} = 0$; i.e. we have $\phi_a \partial^\mu \phi_a = 0, \ (\forall \mu = 0,...,3)$:

For, $m = 1$, the right hand side of Eq.(17) will be

$$v^a_{(1)} \mathcal{K}_a = -\varepsilon_{1il} \varepsilon_{jkl} \left( \frac{3}{2a^3 e} \right) F_{\mu\nu} \phi_i \partial^\mu \phi_j \partial^\nu \phi_k = \left( \frac{3}{a^3 e} \right) \left[ F_{\mu\nu} \left( \phi_2 \partial^\mu \phi_2 + \phi_3 \partial^\mu \phi_3 \right) \partial^\nu \phi_1 \right], \tag{19}$$

and, on account of Eq.(3.a), $\boldsymbol{\phi} \cdot \delta \boldsymbol{\phi} = 0$, it reduces to:

$$v^a_{(1)} \mathcal{K}_a = \frac{-3}{a^3 e} F_{\mu\nu} \phi_1 \partial^\mu \phi_1 \partial^\nu \phi_1 \equiv 0, \tag{20}$$

which vanishes identically since $F_{\mu\nu}$ is anti-symmetric; see Eq.(6).

Similarly, for $m = 2$, and again, on account of, $\boldsymbol{\phi} \cdot \delta \boldsymbol{\phi} = 0$ in Higgs vacuum

$$v^a_{(2)} \mathcal{K}_a = \varepsilon_{2il} \phi_i u^a_{(l)} \mathcal{K}_a = \frac{3}{a^3 e} F_{\mu\nu} \left[ \phi_1 \partial^\mu \phi_1 + \phi_3 \partial^\mu \phi_3 \right] \partial^\nu \phi_2 = \frac{-3}{a^3 e} F_{\mu\nu} \phi_2 \partial^\mu \phi_2 \partial^\nu \phi_2 \equiv 0 \tag{21}$$

Again, it vanishes identically since $F_{\mu\nu}$ is anti-symmetric.

Eqs.(19,20,21), show that, Eq.(3.a), that defines the Higgs vacuum, is a *sufficient* condition for the equations of motion, Eqs.(17), to be satisfied identically there, provided $F_{\mu\nu}$ is anti-symmetric as is the case, here, in the Corrigan-Olive-Fairlie-Nuyts solution, see Eq.(6) above.

To prove the stability of the Higgs vacuum; we need also to show that the condition, $\boldsymbol{\phi} \cdot \delta \boldsymbol{\phi} = 0$, Eq.(3.a), is not only sufficient but is also a "necessary" condition that needs to be satisfied by any small variation of $\boldsymbol{\phi}$, where initially: $\boldsymbol{\phi} \in \mathcal{M}_0$, in order for the equations of motion, Eqs.(17), not to be violated upon the variation of $\boldsymbol{\phi}$. So, assume that we vary $\boldsymbol{\phi}$, already in the Higgs vacuum initially, in an arbitrary way, call it, $\delta' \boldsymbol{\phi}$, such that, in general, it doesn't necessarily satisfy Eq.(3.a); i.e., in general, assume that here we have: $\boldsymbol{\phi} \cdot \delta' \boldsymbol{\phi} \neq 0$.

We need to show that if, $\delta \left( v^a_{(1)} \mathcal{K}_a \right)$, the variation of Eq.(19) due to varying the Higgs field, vanishes, *so as to keep the equations of motion from being violated*; then necessarily: $\boldsymbol{\phi} \cdot \delta' \boldsymbol{\phi} = 0$.

Let: $\phi \to \phi' \equiv \phi + \delta'\phi$, and using Eqs.(19, 20), and that, $\phi$ is in the Higgs vacuum initially,

$$\delta'(v^a_{(1)}\mathcal{K}_a) = \left(\frac{3}{a^3 e}\right)\left[\{\delta' F_{\mu\nu}\}(\phi_2 \partial^\mu \phi_2 + \phi_3 \partial^\mu \phi_3)\partial^\nu \phi_1 + F_{\mu\nu}\{\delta'(\phi_2 \partial^\mu \phi_2 + \phi_3 \partial^\mu \phi_3)\}\partial^\nu \phi_1 \right.$$
$$\left. + F_{\mu\nu}(\phi_2 \partial^\mu \phi_2 + \phi_3 \partial^\mu \phi_3)\delta'\{\partial^\nu \phi_1\}\right] =$$
$$= \left(\frac{3}{a^3 e}\right)\left[-\{\delta' F_{\mu\nu}\}(\phi_1 \partial^\mu \phi_1)\partial^\nu \phi_1 + F_{\mu\nu}\{\delta'(\phi_2 \partial^\mu \phi_2 + \phi_3 \partial^\mu \phi_3)\}\partial^\nu \phi_1 - F_{\mu\nu}(\phi_1 \partial^\mu \phi_1)\delta'\{\partial^\nu \phi_1\}\right]$$

(22)

Where, $\{\delta' F_{\mu\nu}\}(\phi_1 \partial^\mu \phi_1)\partial^\nu \phi_1$, vanishes since, $\delta' F_{\mu\nu}$, is also anti-symmetric.

[Due to the anti-symmetry of $F_{\mu\nu}$]: $F_{\mu\nu}\delta'(\phi_1 \partial^\mu \phi_1 \partial^\nu \phi_1)$ vanishes, and Eq.(22) will reduce to:

$$\delta'(v^a_{(1)}\mathcal{K}_a) = \left(\frac{3}{a^3 e}\right)\left[F_{\mu\nu}\partial^\nu \phi_1\{\delta'(\phi_2 \partial^\mu \phi_2 + \phi_3 \partial^\mu \phi_3)\} + F_{\mu\nu}\partial^\nu \phi_1\{\delta'(\phi_1 \partial^\mu \phi_1)\}\right]$$
$$= \left(\frac{3}{a^3 e}\right)F_{\mu\nu}\partial^\nu \phi_1\{\delta'(\phi_1 \partial^\mu \phi_1 + \phi_2 \partial^\mu \phi_2 + \phi_3 \partial^\mu \phi_3)\}$$

(22.a)

$F_{\mu\nu}\partial^\nu \phi_1$ is, in general, non-vanishing, so:

$$\delta'(v^a_{(1)}\mathcal{K}_a) = 0, \text{ if and only if, } \delta'(\phi_1 \partial^\mu \phi_1 + \phi_2 \partial^\mu \phi_2 + \phi_3 \partial^\mu \phi_3) = 0. \tag{23}$$

Where, same result is true for $\delta'(v^a_{(2)}\mathcal{K}_a) = 0$; starting from Eq.(21).

So, using Eq.(23), and that $\phi$ initially satisfies Eq.(3.a), and keeping only terms up to first order in variation; we have:

$$\phi' \cdot \partial^\mu \phi' \equiv (\phi + \delta'\phi) \cdot \partial^\mu (\phi + \delta'\phi) = \phi \cdot \partial^\mu \phi + \delta'(\phi \cdot \partial^\mu \phi) + \delta'\phi \cdot \partial^\mu \delta'\phi = 0,$$

or, equivalently: $\delta'(\phi \cdot \partial^\mu \phi) = 0 \Leftrightarrow \phi' \cdot \partial^\mu \phi' = 0 \ (\Leftrightarrow \phi \cdot \delta'\phi = 0)$.

(24)

So, [from Eqs.(23, 24)], for any variation, $\phi' \equiv \phi + \delta'\phi$: Equations of motion, Eq.(17), will be violated unless Eq.(3.a) is satisfied. i.e. Eq.(3.a) is a *necessary* condition for Eq.(17) to be satisfied.

## 5. Discussion and Conclusion.

From the results above we note the following:

1) We discussed, above, the stability of the monopole's outer field, i.e. its Higgs vacuum, in the Corrigan-Olive-Fairlie-Nuyts solution by showing explicitly, based on the equations of motion, the restoration of the monopole's Higgs vacuum against any perturbation that violates the Higgs vacuum definition given by Eqs.(2,3). It was shown, that the Higgs vacuum is stable against perturbations in the sense that, the equations of motion, Eqs.(17), do restore the Higgs fields, which are initially in the Higgs vacuum, if perturbed arbitrarily, back to the Higgs vacuum by not permitting them to violate Eq.(3.a), or, to develop, in time, away from the vacuum manifold, $\mathcal{M}_0$; and that was proven by showing that Eq.(3.a) is a sufficient condition that guarantees that Eqs.(17) are being identities of the equations of motion, and inversely, that Eqs.(17) guarantee that the asymptotic Higgs field, which is, initially, in the Higgs vacuum, will not, develop in time off the vacuum manifold, $\mathcal{M}_0$; and this is true since, see Eqs.(23, 24), for any small perturbation of the Higgs fields that violates Eq.(3.a), then $v^a_{(m)}\mathcal{K}_a$, in Eqs.(17), will fail to vanish; i.e., the equations of motion will be violated.

So, due to the equations of motion, the Higgs vacuum remains preserved and get restored asymptotically if perturbed there; and asymptotically, the Higgs field suffers only continuous deformation of the Higgs field in isospace without violating the defining equations of the Higgs vacuum given by Eqs.(2,3).

2) In case at hand, note that the 't Hooft-Polyakov monopole cannot decay while the Higgs vacuum stay preserved and restored by the equations of motion asymptotically when perturbed, since the magnetic charge remains invariant if the Higgs vacuum is preserved asymptotically; and to see that, recall that the magnetic charge for the 't Hooft-Polyakov monopole is given by Eq.(7) which uses, for its evaluation, the Higgs field and its derivatives "only" on the surface of integral which must lie completely in the preserved Higgs vacuum surrounding the core; also recall, see above next to Eq.(7), that the magnetic charge, g, is [10,19] gauge-invariant and also invariant under continuous nonsingular time development of the Higgs field, provided that when g is evaluated, the surface of integral lies completely in the Higgs vacuum as required by Eq.(7); hence *the magnetic charge will be conserved if the Higgs vacuum is preserved asymptotically*, i.e. the monopole is stable and cannot decay due to arbitrary perturbation of the vacuum's Higgs field.

3) Despite that, so far, we only discussed the dynamical stability of the monopole's outer field based on the equations of motion; but we can also discuss the stability of the monopole against perturbations to the inner core by arguing that it is reasonable to assume that the perturbation of the fields inside the core, will never cause the monopole to decay, i.e. it is stable, since as long as the Higgs vacuum stays preserved asymptotically, which is the case here, then the magnetic charge will be conserved too since it is measured, Eq.(7), based only on Higgs field and its derivatives on the surface of integral which lies completely in the Higgs vacuum region surrounding the core.

4) When studying stability of the monopole due to perturbation of the core, we can avoid studying the complex equations of motion inside the core by simply considering it as a black box as was done by Brandt and Neri [17], see also Coleman [16], on the stability of monopoles. In their classic work [17], in order to verify whether GNO monopole's dynamical solutions are stable, they applied perturbations to the outer asymptotic region outside the monopole's core considering the inner core as a "black box" whose fields are of some smooth structure; their standard way in order to verify the instability of all dynamical GNO monopoles solutions *except for only one GNO solution for each distinct topological class of Lubkin construction* [18], was to verify that the small arbitrary perturbations of the gauge fields in the asymptotic region outside the core will grow exponentially for gauge fields associated with off-diagonal elements in the symmetry group's Lie algebra; this means that, for the unstable solutions, and upon perturbation we will no longer be able to reduce the Lie-algebra-evaluated gauge fields to the GNO form [11] which is the most general solution of the monopole's source-less Yang-Mills-Higgs equations outside the inner core. Recall that the GNO general solution can always be written, using some appropriate gauge transformation, as a diagonal constant element of the Lie algebra multiplied by Dirac's Abelian monopole's field. Failing, upon perturbation, to put the asymptotic monopole's field for the monopole's unstable solutions in the GNO form for a single monopole by any gauge transformation was interpreted [17] as, in the words of Coleman [16], the decay of the monopole by emission of non-Abelian magnetic radiation.

Back now to our case at hand; since we already have shown that, in the Corrigan-Olive-Fairlie-Nuyts solution, the 't Hooft-Polyakov monopole's Higgs vacuum will always remain asymptotically preserved, due to the equations of motion, against any perturbations. But we already know that, see Eq(5) above, in the Higgs vacuum we have:

$$\mathbf{G}^{\mu\nu} = \frac{1}{a} \boldsymbol{\phi} F^{\mu\nu}; \qquad (5')$$

and this means that, even if monopole is perturbed, the only the field strength surviving asymptotically outside the monopole's core will be the field strength associated with $\boldsymbol{\phi}$'s isospin direction; while the field strengths, associated with the other isospin directions orthogonal to $\boldsymbol{\phi}$'s, will remain confined to the monopole's inner core and corresponds to the "massive" gauge fields of short distance range inside the core; *with gauge fields masses*

*resulting from Higgs mechanism.* This means that as long as the Higgs vacuum is preserved asymptotically, as was shown above to be case, then Eq (5′) must be satisfied allowing only Abelian field asymptotically, and hence the perturbation of the massive-short-ranged gauge fields inside the inner core can only cause the massive fields' fluctuation about ground state inside the core; and the core can recover its ground state only by emitting Abelian radiation asymptotically since the Higgs vacuum will have to survive in the asymptotic region outside the monopole's core at all times as required by the equations of motion; that means that the non-Ablian radiation is not allowed asymptotically; i.e., massive gauge fields will have to be confined to the inner core which will be intact, and that shows stability of monopole against the inner core perturbation too.

Note also that the survival of the monopole's Higgs vacuum asymptotically, as required by the equations of motion, will also guarantee that the monopole's system will maintain its energy finiteness and thence adds to the verification of the physicality of the Corrigan-Olive-Fairlie-Nuyts solution.

**To summarize:** we considered the general solution given by Corrigan, Olive, Fairlie and Nuyts [12]; we derived constraints from the equations of motion that proved preservation of Higgs vacuum against perturbation. We finally concluded the *monopole's stability*, since the Higgs vacuum, which was shown to be preserved, allows only Abelian radiation asymptotically, as seen from Eq.(5′); and thence the core is not allowed asymptotically to emit non-Abelian radiation, in isospin directions orthogonal to isospin direction of $\phi$, which is necessary if monopole is to decay.